\documentclass[10pt]{article}

\usepackage{cite}
\usepackage{amsmath,amssymb,amsfonts}
\usepackage{graphicx}
\usepackage{caption}
\usepackage{subcaption}
\usepackage{booktabs}
\usepackage{multirow}

\usepackage{tikz}
\usetikzlibrary{shapes,arrows,positioning,matrix,plotmarks}
\usepackage{url}
\usepackage{hyperref}

\title{Adaptive Fault Detection exploiting Redundancy
  with Uncertainties in Space and Time
}

\author{%
  Denise~Ratasich,
  Michael~Platzer,
  Radu~Grosu,
  Ezio~Bartocci\\
  Institute of Computer Engineering,
  TU~Wien, Vienna, Austria\\
  (e-mail: \{firstname.lastname\}@tuwien.ac.at)
}
\date{2019-03-11}

\DeclareMathAlphabet{\mathbfsf}{\encodingdefault}{\sfdefault}{bx}{n}

\newcommand{\var}{\mathsf{v}}

\newcommand{\varp}{\ensuremath{\lowercase{v}}}

\newcommand{\rel}{\mathsf{r}}

\newcommand{\vars}{\ensuremath{\text{\sffamily\uppercase{V}}}}

\newcommand{\rels}{\ensuremath{\text{\sffamily\uppercase{R}}}}

\newcommand{\prior}[1]{\expandafter\hat#1^{(-)}}
\newcommand{\est}[1]{\expandafter\hat#1}

\newcommand{\vinterval}[1]{\ensuremath{\boldsymbol{#1}}}
\newcommand{\vlo}[1]{\ensuremath{\underline{#1}}}
\newcommand{\vhi}[1]{\ensuremath{\overline{#1}}}

\usepackage{listings}

\usepackage{caption}
\DeclareCaptionFormat{sourcefile}{\noindent\colorbox[rgb]{1,0.9,0.4}{\parbox{\dimexpr\textwidth-2\fboxsep}{#3}}}

\usepackage{color}
\definecolor{sourcefilebg}{rgb}{1,1,0.8}
\definecolor{bashbg}{rgb}{0.95,0.95,0.95}
\definecolor{tuwBlue}{RGB}{0,116,178}
\definecolor{tuwGray}{RGB}{102,102,102}

\lstdefinestyle{problog}{
  language=Prolog,
  basicstyle=\ttfamily\footnotesize,
  captionpos=b,
  frame=single, framerule=0pt,
  framesep=5pt, xleftmargin=5pt, xrightmargin=5pt,
  numbers=left, numbersep=10pt, numberstyle=\tiny\color{tuwGray},
  backgroundcolor=\color{sourcefilebg},
  showstringspaces=false,
  escapeinside={(*@}{@*)},
  keywordstyle=\bfseries\color{tuwBlue},
  commentstyle=\itshape\color{tuwGray},
  stringstyle=\color{tuwGray}
}

\lstdefinestyle{bash}{
  language=bash,
  basicstyle=\ttfamily\footnotesize,
  frame=single, framerule=0pt,
  framesep=5pt, xleftmargin=5pt, xrightmargin=5pt,
  backgroundcolor=\color{bashbg},
  keywordstyle=\mdseries\color{black},
  commentstyle=\itshape\color{tuwGray},
  stringstyle=\color{black}
}

\begin{document}

\maketitle

\begin{abstract}
  \textbf{Abstract.}
The Internet of Things (IoT) connects millions of devices
of different cyber-physical systems (CPSs)
providing the CPSs additional (implicit) redundancy during runtime.
However, the increasing level of dynamicity, heterogeneity, and complexity
adds to the system's vulnerability, and challenges its ability to react to faults.
Self-healing is an increasingly popular approach for ensuring resilience, that is,
a proper monitoring and recovery, in CPSs.
This work encodes and searches an adaptive knowledge base in Prolog/ProbLog
that models relations among system variables
given that certain implicit redundancy exists in the system.
We exploit the redundancy represented in our knowledge base
to generate adaptive runtime monitors which compares related signals
by considering uncertainties in space and time.
This enables the comparison of uncertain, asynchronous, multi-rate and delayed measurements.
The monitor is used to trigger the recovery process of a self-healing mechanism.
We demonstrate our approach by deploying it
in a real-world CPS prototype of a rover
whose sensors are susceptible to failure.

\end{abstract}


\section{Introduction}

Cyber-physical systems (CPSs) are desired to be resilient
(i.e., dependable and secure) against different kinds of faults
throughout its lifecycle.
%
Failure scenarios like communication crashes and dead batteries (fail-silent, fail-stop)
are easy to handle (watchdog/timeout).
However, sensor data may be erroneous due to noise (e.g., communication line, aging),
environmental influences (e.g., dirtying, weather)
or a security breach.
To detect a sensor failure one has to
define a valid signature%
~\cite{chandola2009anomaly,butun2014survey,mitchell2014survey,buczak2016survey}
or specification%
~\cite{bartocci2018lectures,falcone2018runtime}
or create particular failure models%
~\cite{kalman1960new,bellanger2001adaptive}
for each possible hazard (c.f., aging, dirtying and a security breach).
A simple method for detecting a faulty sensor value in different failure scenarios
is to check against related information sources, i.e., exploit redundancy.
Explicit redundancy, that is replicating the sensor~\cite{kopetz2011realtime,poledna1996faulttolerant} is costly.

A designed system might not incorporate (implicit) redundancy straightaway
(if so, the system would use information fusion to exploit the redundancy).
However, when assembling (sub-)systems of different suppliers
redundancy often becomes available.
This is typically the case for distributed systems like the Internet of Things (IoT) or a Cyber-Physical System (CPS).
For instance, the drivetrain of an electric vehicle is equipped with several sensors
at the battery, motor, transmission, shaft, differential, or wheels level.
The physical entities (CPS variables) observed in these subsystems,
e.g., power consumption, revolution, speed or torque,
are all related to the velocity of the vehicle,
thus providing implicit information redundancy~%
\cite{hoftberger2015knowledgebased,amorim2018runtime}.

We exploit such implicit redundancy
by encoding it in a knowledge base~\cite{hoftberger2015knowledgebased,ratasich2018selfhealing}
created by domain experts or learned during runtime.
Our previous work~\cite{ratasich2017selfhealing}
shows how to use the redundancy to substitute failed observation components.
We refer to it as \emph{Self-Healing by Structural Adaptation} (SHSA).
The SHSA service acts independently of the application on the communication network of the system.
However, it needs a monitor to trigger the recovery from failures.

This work exploits the SHSA knowledge base to monitor related information.
Similar solutions~\cite{papa2012transfer,ntalampiras2015detection,zhang2018fault}
perform the comparison synchronously, that is,
all measurements are assumed to have the same timestamp
(and therefore can be safely compared against each other).
However, typically the runtime monitor has to cope with
asynchronous, multi-rate and delayed measurements.
Moreover, the timestamp of a measurement might not express the time
when the CPS variable actually adopted this value (time-delay system).
Typical runtime monitors are statically configured
and therefore cannot cope with changes of a dynamic and evolving system
(e.g., availability of signals, message structure of the observations).

We therefore propose an observation model based on interval arithmetic
considering uncertainties in space \emph{and time}
and an adaptive method to compare such signals.
In particular, our novel contributions are as follows:
\begin{itemize}
\item A short, adaptive and extensible implementation of the knowledge base in Prolog/ProbLog.
\item A simple model for observations of a signal that is prone to uncertainties in space and time.
\item The generation of adaptive runtime monitors for signals
  which compare uncertain, asynchronous, multi-rate and/or delayed observations.
\item Demonstration of our method on a real-world application
  -- an implementation on a rover prototype
  that performs collision avoidance.
\end{itemize}

The rest of the paper is organized as follows.
Section~\ref{sec:relatedwork} gives an overview of related work.
Section~\ref{sec:background} introduces the running example of our mobile robot
and defines the knowledge base.
Section~\ref{sec:prolog} encodes the knowledge base in Prolog.
Section~\ref{sec:monitor} describes the signal model
and shows how a monitor is generated for comparing itoms.
Section~\ref{sec:experiments} demonstrates and tests
the fault detection of the generated monitor.
Section~\ref{sec:conclusion} concludes by summarizing our results
and outlining the future work.


\section{Related Work}
\label{sec:relatedwork}

In~\cite{hoftberger2015knowledgebased}, H{\"o}ftberger introduced 
an ontology (a knowledge base) defining physical relations or semantic 
equivalences between variables (e.g., laws of physics)
to substitute failed observation services.
We applied this technique in~\cite{ratasich2017selfhealing}
and we extended the knowledge base with properties and utility 
theory in~\cite{ratasich2018selfhealing}.
However, this knowledge base has been only exploited for 
failure recovery, but not yet for fault detection.

Literature on fault detection is spread to the area of
runtime verification~\cite{bartocci2018lectures,falcone2018runtime},
anomaly detection~\cite{chandola2009anomaly},
intrusion detection~\cite{butun2014survey,mitchell2014survey,buczak2016survey},
fault-tolerance~\cite{avizienis2004basic,sathya2010survey,kopetz2011realtime,herault2015faulttolerance}
and self-healing~\cite{ghosh2007selfhealing,psaier2011survey}.
Approaches to detect faults typically
use a specification or signature, an anomaly model or redundancy as a reference
to judge the behavior of an entity.
%
%

A popular approach to perform fault detection in safety 
critical systems is to exploit redundancy of components. 
Examples include lockstep~\cite{poledna1996faulttolerant} and triple modular 
redundancy (TMR)~\cite{kopetz2011realtime} that
explicitly replicate software or hardware components to detect faults.
In the best case the replicas are diverse,
i.e., they have been separately designed but share the same functionality and interface,
to avoid joint or concurrent failures.
Replication is generally expensive because it requires 
additional time and resources both at design time and runtime.

The security framework described in~\cite{durrwang2017security}
performs anomaly detection on input signals of a component in a distributed network
by comparing the input against trusted local signals (if available).
The local signals may be provided by sensors
which are directly sampled by the component
and not advertised in the vulnerable communication network (cf. IoT) of the system.

The proposed method in~\cite{papa2012transfer}
votes over signals sent through transfer functions (cf. relations)
of known or unknown relationships of signals.
The difference between the signals raises an alert
when it exceeds a threshold.
Similarly, we use transfer functions (here: relations)
to bring signals into a common domain before comparison.

The author of~\cite{ntalampiras2015detection}
uses ensemble modeling to represent the redundancy of data streams.
First a dependency graph between data streams of nodes is created (via cross-correlation).
An ensemble of estimators of the learned models is defined
given a threshold on the cross-correlation.
The estimators can be created from different types of models
(e.g., linear functions or neural networks)
to relate the data streams to each other
(temporal and functional relationships in the physical layer).
The output of the estimators are combined in a weighted-averaging fashion
(cf. confidence-weighted averaging~\cite{elmenreich2007fusion}).
Another threshold for the estimator output defines whether an anomaly has occurred or not.
However, the ensemble is statically defined during design time.
The estimators have to be aggregated and weighted in a proper way.
Similarly, the threshold is crucial to avoid false positives.
Although time delays are not discussed,
appropriate estimators can be set up to consider (at least constant) time delays.

The authors of~\cite{zhang2018threshold}
propose instead a Bayesian network (BN) with past and future values of sensor readings and redundant ones
(spatial and temporal redundancy)
to estimate the true state of a variable.
A sensor is classified faulty, when its measurement exceeds a threshold given the estimated state.
Similarly, fault injection is used to evaluate the monitoring technique.
The proposed BN can be compared to a state estimator fusing redundant sensors.

\newlength{\sotatabwidth}
\sotatabwidth0.17\linewidth
\begin{table}[h]
  \centering
  \resizebox{\linewidth}{!}{%
  \begin{tabular}{|p{\sotatabwidth}|p{0.7\sotatabwidth}|p{0.7\sotatabwidth}|p{0.7\sotatabwidth}|p{0.7\sotatabwidth}|p{0.9\sotatabwidth}|}
    \hline
    \multirow{2}{\sotatabwidth}{\textsc{References}}
    & \multicolumn{2}{c|}{\textsc{Redundancy}}
    & \multirow{2}{\sotatabwidth}{\textsc{Dynamic Systems}}
    & \multicolumn{2}{c|}{\textsc{Uncertain Signals}}
    \\ \cline{2-3} \cline{5-6}
    & \textsc{Explicit} & \textsc{Implicit} &
    & \textsc{Space} & \textsc{Time}
    \\ \hline
    \hline
    \cite{poledna1996faulttolerant,kopetz2011realtime} & \checkmark & & & & \\ \hline
    \cite{durrwang2017security} & \checkmark & \checkmark & & & \\ \hline
    \cite{papa2012transfer} & & \checkmark & & & \\ \hline
    \cite{zhang2018threshold} & \checkmark & & & \checkmark & \\ \hline
    \cite{ntalampiras2015detection} & \checkmark & \checkmark & & \checkmark & \\ \hline
    This work & \checkmark & \checkmark & \checkmark & \checkmark & \checkmark \\ \hline
  \end{tabular}%
  }
  \caption{Comparison of state-of-the-art fault detection using redundancy.}
  \label{tab:sota}
\end{table}

All related work listed in Table~\ref{tab:sota} considers constant variance of the signals,
i.e., the threshold for mismatch is set a priori.
By using intervals to express observations we can handle time-varying errors.
Most of the related work assumes the signals to compare
to be sampled at the same point in time, i.e., are truly comparable.
A dynamic BN like proposed by~\cite{zhang2018threshold} could relate asynchronous observations
to each other by forward estimation given the timestamp of the observation.
However, neither of the listed work discussed this issue.
Our knowledge base resembles a Bayesian network (in particular the sensor model of a BN),
similar to \cite{ntalampiras2015detection,zhang2018threshold},
but with deterministic nodes (relations or transfer functions) that can be re-used, merged and adapted
when components are added or removed from a dynamic system.
However, for our needs the knowledge base expresses a set of rules (cf. relations)
on how to compare signals or to substitute a failed signal.


\section{Background}
\label{sec:background}

In this section we recap our knowledge base of redundancy
and showcase it on a real-world prototype~\cite{ratasich2018selfhealing}.


\subsection{Running Example}

\begin{figure}[t]
  \centering
  \includegraphics[width=\linewidth]{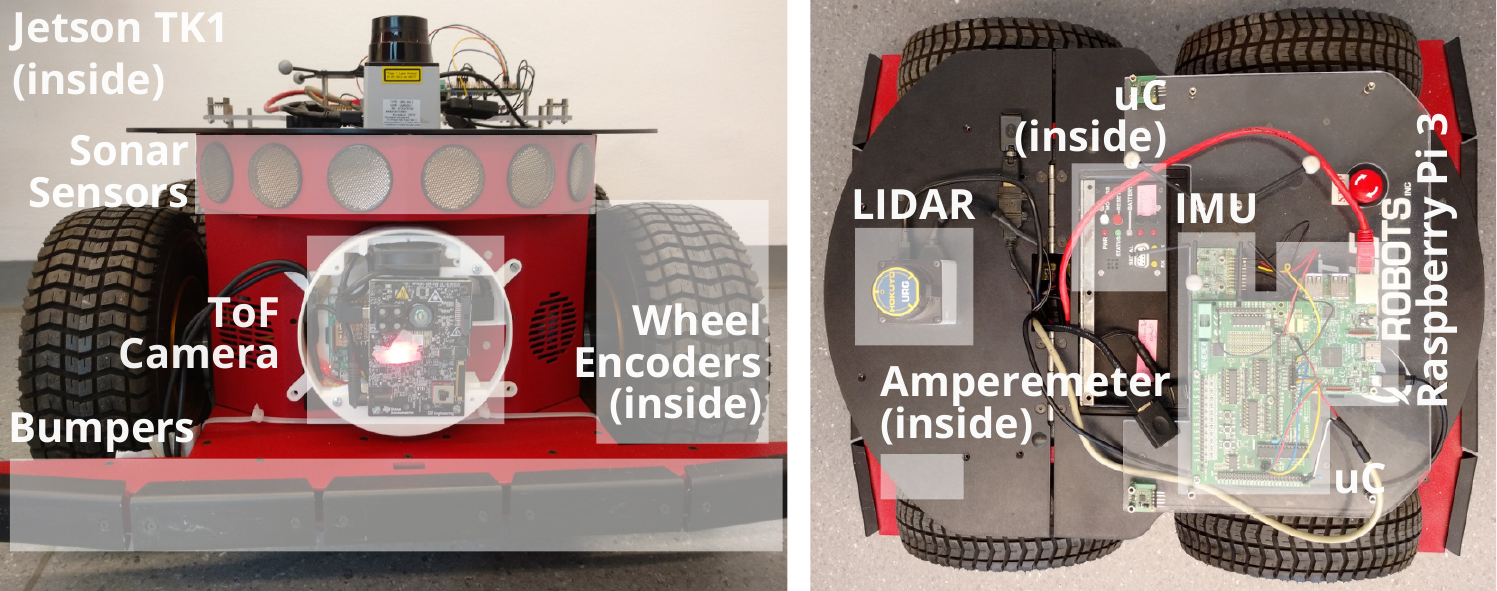}
  \caption{Mobile robot equipped with its sensors and processing units.}
  \label{fig:robot}
\end{figure}

The running example is a mobile robot (Pioneer 3-AT) that avoids collisions.
Our rover under test~(Fig.~\ref{fig:robot}) is equipped with a Jetson TK1 and a Raspberry~Pi~3
running the Robot Operating System~(ROS)~\cite{quigley2009ros} and controlled 
via WiFi using a notebook. A ROS application can be distributed into several 
processes, so-called \emph{nodes} which may run on different machines. Nodes 
communicate via a message-based interface over TCP/IP. In particular, ROS nodes 
subscribe and publish to ROS topics (cf. named channels). ROS can start new 
nodes and reconfigure the communication flow of existing nodes during runtime 
and it is therefore suitable for SHSA~\cite{ratasich2017selfhealing}.

\subsection{System Model}
As our focus is on self-healing in the software cyber-part of a CPS $Z$
(cf. dynamic reconfiguration of an FPGA),
we assume that each physical component comprises at least one software component $z$
(e.g., the driver of the LIDAR)
and henceforth, consider the software components only.

A system $Z$ can be characterized by properties, referred to as system features,
or simply as \emph{variables}~$\vars$
(e.g., position of a vehicle, point cloud or distance measurements acquired by a LIDAR).
In SHSA we focus on observations of CPS variables and
refer to the actual data as \emph{value}.
A \emph{signal} $\varp(t)$ or short $\varp$ represents the values of a variable $\var$ over time ($t \in \mathbb{R}$)
provided by a certain component (e.g., a sensor).

The value of a system variable -- an observation -- is communicated between components
typically via message-based interfaces.
We denote such transmitted data representing the value of a variable~$\var$,
as information atom~\cite{kopetz2014conceptual}, short \emph{itom}.
In general, an itom is a tuple including data and an explanation of this data.
An SHSA itom $\varp(t_s) = (v,t_s)$ is a sample or a snapshot of a variable at a certain point in time $t_s$.
It provides the following explanation of the data:
\emph{i)}~the associated variable or entity, and
\emph{ii)}~the timestamp of the acquisition or generation of the value.
Each itom is identified by its signal's name $\varp$ and its timestamp.
A variable can be provided by different components simultaneously
(e.g., $p$ redundant sensors).

Each software component $z$ executes a program that uses input signals
and provides output signals.

The CPS implements some functionality, a desired service (e.g., tracking).
The subset of components implementing the CPS' objectives are called
\emph{controllers}.

A signal~$\varp$ is \emph{needed}, when $\varp$ is input of a controller.

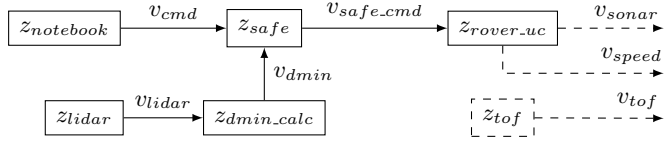
\begin{figure}[t]
  \centering
\tikzstyle{component} = [draw, rectangle, inner sep = 4pt]
\tikzstyle{unused} = [dashed]

\begin{tikzpicture}[auto, >=latex, font=\sffamily\small]

  \node [component] (z1) {$z_{lidar}$};
  \node [component, right = 3em of z1] (z2) {$z_{dmin\_calc}$};
  \node [component, above = 2em of z2] (z4) {$z_{safe}$};
  \node [component, left = 4em of z4] (z3) {$z_{notebook}$};
  \node [component, right = 5.5em of z4] (z5) {$z_{rover\_uc}$};
  \node [right = 4em of z5] (n1) {};
  \node [below = 1em of n1] (n2) {};
  \node [component, below = 2em of z5, unused] (z6) {$z_{tof}$};
  \node [below = 1em of n2] (n3) {};

  \draw [->] (z1) -- (z2) node [pos=0.5] {$\varp_{lidar}$};
  \draw [->] (z2) -- (z4) node [pos=0.5,right] {$\varp_{dmin}$};
  \draw [->] (z3) -- (z4) node [pos=0.5] {$\varp_{cmd}$};
  \draw [->] (z4) -- (z5) node [pos=0.5] {$\varp_{safe\_cmd}$};
  \draw [->, unused] (z5) -- (n1) node [pos=0.65] {$\varp_{sonar}$};
  \draw [->, unused] (z5) |- (n2) node [pos=0.9] {$\varp_{speed}$};
  \draw [->, unused] (z6) -- (n3) node [pos=0.8] {$\varp_{tof}$};
\end{tikzpicture}

  \caption{Relevant application components and signals of the mobile robot.}
  \label{fig:application-nodes}
\end{figure}

\textbf{Case Study} (Fig.~\ref{fig:application-nodes}):
The rover is tele-operated by the controller $z_{notebook}$ publishing the
desired velocity $\varp_{cmd}$.
The distance to the nearest obstacle $\var_{dmin}$ is evaluated
by the component $z_{dmin\_calc}$ using the LIDAR data as input.
As soon as $\varp_{dmin}$ falls below a safety margin (for simplicity a constant value)
the robot's verified desired velocity $\varp_{safe\_cmd}$ is set to $0$.
The component $z_{rover\_uc}$ applies $\varp_{safe\_cmd}$ to the motors
and provides sonar range measurements $\varp_{sonar}$
and actual linear and angular velocity $\varp_{speed}$.

Another component $z_{tof}$ provides 3D images of the environment in front of the rover.
However, dashed signals in Fig.~\ref{fig:application-nodes}
are not used for collision avoidance but available for other tasks 
(e.g., parking, object recognition).

\subsection{Input Interface of Components}

The messages communicated within the network may be received synchronously or asynchronously.
In our rover like in many other IoT infrastructures, the communication is asynchronous,
i.e., the exact point in time of a message reception is unknown.
The task of the component might be executed when an itom is received
(e.g., $z_{dmin\_calc}$ calculating minimum distance when receiving $\varp_{lidar}$).
However, the inputs of a software component might be published with different rates.
Some components therefore have to cope with multi-rate, asynchronous and late messages.
The monitor described in Sec.~\ref{sec:monitor} periodically executes the plausibility check
using the itoms collected since the last monitor call
and a (configurable sized) buffer of itoms of previous executions.

\subsection{Knowledge Base of Redundancy}
\label{sec:knowledgebase}

This section defines the knowledge base used to describe implicit redundancy~%
\cite{ratasich2018selfhealing}.

\subsubsection{Variables and Relations}

Variables are related to each other.
A relation $\rel : \var_o = f(\vars_I)$ is a function or program
(e.g., math, pseudo code or executable python code)
to compute an output variable~$\var_o$
from a set of input variables~$\vars_I$.
The relations can be defined by the application's domain expert
or learned (approximated) with neural networks, SVMs or polynomial functions
(see \cite{ratasich2017selfhealing}).

\subsubsection{Structure}
\label{sec:knowledgebase_structure}

The knowledge base $K = (\vars, \rels, E)$ is a bipartite directed graph
(which may also contain cycles)
with independent sets of variables $\vars$ and of relations $\rels$ of a CPS.
$\vars$ and $\rels$ are the nodes of the graph.
Edges $E$ specify the input/output interface of a relation.
In particular, $\var_i$ is an input variable for $\rel$
iff $\exists (\var_i, \rel) \in E$
denoted as $\var_i \xrightarrow{e} \rel$.
$\var_o$ is the output variable of $\rel$
iff $\exists (\rel, \var_o) \in E$
denoted as $\rel \xrightarrow{e} \var_o$.
$Pred_Y(x)$ denotes the predecessors of a node $x$ in graph $Y$.
There are no bidirectional edges, i.e., if
$\var_i \xrightarrow{e} \rel \Rightarrow {\not\exists} \rel \xrightarrow{e} \var_i$.
Hence a variable is either input or output to a relation, but never both.
A relation can further have only one output variable, i.e., for
$\forall j \neq i$ if $\rel \xrightarrow{e} \var_i \Rightarrow {\not\exists} \rel \xrightarrow{e} \var_j$.
Note that relations have to be modeled as nodes, not edges,
because a variable is typically related to more than one variable via a single relation.

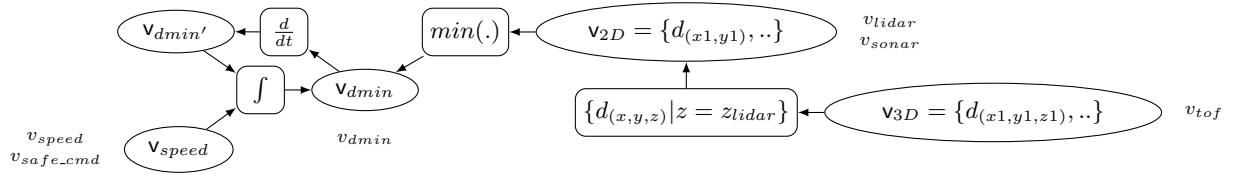
\begin{figure*}[htb]
  \centering

\tikzstyle{var} = [draw, ellipse, inner sep=3pt, align=center]
\tikzstyle{rel} = [draw, rectangle, rounded corners, inner sep=3pt, minimum size = 1.8em]
\tikzstyle{itom} = [rectangle, align=center, font=\scriptsize]

\begin{tikzpicture}[auto, node distance=1em, >=latex, font=\sffamily\small]

  \node [var] (v1) {$\var_{dmin}$};
  \node [itom, below = 0.5em of v1] (i1) {
    $\varp_{dmin}$%
  };

  \node [rel, left = of v1] (r1) {$\int$};
  \node [var, above left = of r1] (v11) {$\var_{dmin'}$};
  \node [var, below left = of r1] (v12) {$\var_{speed}$};
  \node [itom, left = 0.5em of v12] (i12) {
    $\varp_{speed}$\\
    $\varp_{safe\_cmd}$%
  };

  \draw [->] (v11) -- (r1);
  \draw [->] (v12) -- (r1);
  \draw [->] (r1) -- (v1);

  \node [rel, above left = of v1] (rd) {$\frac{d}{dt}$};

  \draw [->] (v1) -- (rd);
  \draw [->] (rd) -- (v11);

  \node [rel, above right = of v1] (r2) {$min(.)$};
  \node [var, right = of r2] (v21) {$\var_{2D} = \{d_{(x1,y1)}, ..\}$};
  \node [itom, right = 0.5em of v21] (i21) {
    $\varp_{lidar}$\\
    $\varp_{sonar}$%
  };

  \draw [->] (v21) -- (r2);
  \draw [->] (r2) -- (v1);

  \node [rel, below = of v21] (r3) {$\{ d_{(x,y,z)} | z = z_{lidar} \}$};
  \node [var, right = of r3] (v31) {$\var_{3D} = \{d_{(x1,y1,z1)}, ..\}$};
  \node [itom, right = 0.5em of v31] (i31) {
    $\varp_{tof}$%
  };

  \draw [->] (v31) -- (r3);
  \draw [->] (r3) -- (v21);

\end{tikzpicture}
  \caption{Knowledge base of the mobile robot
    depicting relations exploitable for collision avoidance.
    Boxes are relations.
    Ellipses are variables.
    Variables are annotated with available signals (from Figure~\ref{fig:application-nodes}).
  }
  \label{fig:knowledge-base}
\end{figure*}

\textbf{Case Study}:
The application nodes of the rover (Fig.~\ref{fig:application-nodes})
provide several observation signals.
The relations of the signals, that is the knowledge base, are set up manually (Fig.~\ref{fig:knowledge-base}).

\subsubsection{Substitution}

A substitution $s$ of $\var_{sink}$ is a connected acyclic sub-graph
of the knowledge base with the following properties:
\emph{i)}~The output variable is the only sink of the substitution
(acyclic + Eq.~\ref{eq:onesink}).
\emph{ii)}~Each variable has zero or one relation as predecessor
(Eq.~\ref{eq:onerelation}).
\emph{iii)}~All input variables of a relation must be included
(Eq.~\ref{eq:sources}; it follows that the sources of the substitution graph are variables only).

\begin{align}
    &s = (\vars_s, \rels_s, E_s) \\
    &\var_{sink} \in \vars_s, \vars_s \subseteq \vars, \rels_s \subseteq \rels, E_s \subseteq E \nonumber\\
    \label{eq:onesink}
    &\forall x \in \rels_s \cup \vars_s \setminus \var_{sink} \quad
      \exists y \in
      \rels_s \cup \vars_s \; | \; \exists x \xrightarrow{e \in E_s} y \\
    \label{eq:onerelation}
    &\forall \var \in \vars_s \quad |Pred_s(\var)| \le 1\\
    \label{eq:sources}
    &\forall \rel \in \rels_s \quad \forall \var \in Pred_K(\rel) \; | \; \var \in \vars_s
  \end{align}

A substitution $s$ is \emph{valid} if all sources are provided,
otherwise the substitution is \emph{invalid}.
We denote the set of valid substitutions of a variable $\var$ as $S(\var)$.
Only a valid substitution can be instantiated
by concatenating the relations $\rels_s$ to the function~$f_s$
which takes selected signals~$I_s$ as input.

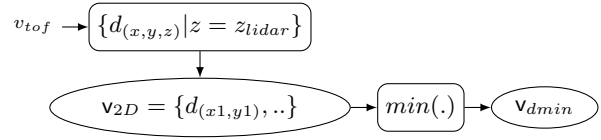
\begin{figure}[htb]
  \centering

\tikzstyle{var} = [draw, ellipse, inner sep=3pt, align=center]
\tikzstyle{rel} = [draw, rectangle, rounded corners, inner sep=3pt, minimum size = 1.8em]
\tikzstyle{itom} = [rectangle, align=center, font=\scriptsize]

\begin{tikzpicture}[auto, node distance=1em, >=latex, font=\sffamily\small]

  \node [var] (v1) {$\var_{dmin}$};

  \node [rel, left = of v1] (r2) {$min(.)$};
  \node [var, left = of r2] (v21) {$\var_{2D} = \{d_{(x1,y1)}, ..\}$};

  \draw [->] (v21) -- (r2);
  \draw [->] (r2) -- (v1);

  \node [rel, above = of v21] (r3) {$\{ d_{(x,y,z)} | z = z_{lidar} \}$};
  \node [itom, left = of r3] (v31) {$\varp_{tof}$};

  \draw [->] (v31) -- (r3);
  \draw [->] (r3) -- (v21);

\end{tikzpicture}
  \caption{A valid substitution for $\var_{dmin}$.}
  \label{fig:substitution}
\end{figure}

\textbf{Case Study}:
When $\varp_{dmin}$ fails it can be replaced
by the outputs of the substitution in Figure~\ref{fig:substitution},
by using the provided signal $\varp_{tof}$ as input.


\section{SHSA in Prolog}
\label{sec:prolog}

A Prolog program is a set of facts, rules and queries.
The formal definitions of SHSA given above
can be efficiently (that is, with a few lines of code) expressed in Prolog.
It is also a good basis to extend SHSA in future work, e.g.,
runtime adaptation of the knowledge base or
validation of requirements for substitutions.

The structure of the knowledge base and availability of variables
is defined by a set of Prolog facts (see example in Listing~\ref{lst:kb}).
In particular, a relation $\rel : \var_o = f(\vars_I)$
is written as the fact \texttt{function(o,r,[i0,..,in])}
with a list of names \texttt{[i0,..,in]} representing the input variables,
a name of the relation \texttt{r}
and the name of the output variable \texttt{o}.
An itom \texttt{i} is available when the fact \texttt{itom(i)} evaluates to true.
In addition, each itom has to be mapped to its corresponding variable.
For convenience, the availability of a variable can be specified via a list of itoms, e.g.,
\texttt{itomOf(v,[i0,..,in])}.
This implicitly maps each itom \texttt{i*} to the variable \texttt{v}
and marks the variable \texttt{v} as \texttt{provided(v)}
(\emph{provided} is a Prolog property).
The facts \texttt{itomsOf(.)}
are added during runtime to the program (structure of the knowledge base)
depending on the itoms gathered by the monitor.

\textbf{Case Study}:
Listing~\ref{lst:kb} shows the knowledge base of Figure~\ref{fig:knowledge-base} in Prolog.
Table~\ref{tab:name-mapping} maps the knowledge base variables to Prolog names.

\begin{lstlisting}[style=problog,
    caption={Knowledge base of the mobile robot in Prolog.},
    label=lst:kb]
:- use_module(library(shsa)).

% structure
function(dmin, r1, [d_2d]).
function(d_2d, r2, [d_3d]).
function(dmin, r3, [dmin_last, speed]).
function(dmin_last, r4, [dmin]).

% provided itoms
itomsOf(dmin, ["/emergency_stop/dmin/data"]).
itomsOf(d_2d, ["/p2os/sonar/ranges",
               "/scan/ranges"]).
itomsOf(d_3d, ["/tof_camera/frame/depth"]).
itomsOf(speed, ["/p2os/cmd_vel",
                "/p2os/odom"]).
\end{lstlisting}

\begin{table}[h]
  \centering
  \begin{tabular}{|l|l|l|l|}
    \hline
    \textsc{Variable} & \textsc{Prolog} & \textsc{Signal} & \textsc{Prolog} \\ \hline
    \hline
    $\var_{dmin}$ & \verb|dmin|
    & $\varp_{dmin}$ & \verb|"/calculator/dmin"| \\ \hline
    $\var_{dmin'}$ & \verb|dmin_last| & - & - \\ \hline
    $\var_{speed}$ & \verb|speed|
    & $\varp_{speed}$ & \verb|"/rover/act_vel"| \\ \cline{3-4}
    & & $\varp_{safe\_cmd}$ & \verb|"/safe/cmd_vel"| \\ \hline
    $\var_{2D}$ & \verb|d_2d|
    & $\varp_{sonar}$ & \verb|"/sonar/ranges"| \\ \cline{3-4}
    & & $\varp_{lidar}$ & \verb|"/lidar/ranges"| \\ \hline
    $\var_{3D}$ & \verb|d_3d|
    & $\varp_{tof}$ & \verb|"/tof/ranges"|\\ \hline
  \end{tabular}%
  \caption{Used variable and signal names in Prolog.}
  \label{tab:name-mapping}
\end{table}

The Prolog library \texttt{shsa} implements a set of rules to
evaluate properties (e.g., SHSA properties of variables like \texttt{provided(v).})
or to search substitutions for a needed variable (query \texttt{substitution(v,S).}).
It basically implements the definitions in the previous section
within 22 lines of code. Details can be found in our repository~%
\footnote{\url{https://github.com/dratasich/shsa-prolog}}.

Finally, we can query properties, e.g., \texttt{substitution(v,S).}
to find all valid substitutions for the variable \texttt{v}.
The Prolog engine (here: ProbLog\cite{dries2015problog2})
tries to satisfy the query, that is,
it searches for \texttt{S} which evaluates the query to true (``unify''~\texttt{S})
considering the given facts (SHSA knowledge base) and rules (SHSA library).
In this case, \texttt{S} is a relation or a tree of relations,
see the substitute search for \texttt{dmin} in Listing~\ref{lst:query}.

\begin{lstlisting}[style=problog,
    caption={Evaluation result of the query \texttt{substitution(dmin,S)}.},
    label=lst:query]
substitution(dmin,"/emergency_stop/dmin/data")
substitution(dmin,[function(dmin,r1,[d_2d]),
                   "/scan/ranges"])
substitution(dmin,[function(dmin,r1,[d_2d]),
                   "/p2os/sonar/ranges"])
substitution(dmin,[function(dmin,r1,[d_2d]),
                   [function(d_2d,r2,[d_3d]),
                    "/tof_camera/frame/depth"]])
\end{lstlisting}
The complexity of the query is the one of a depth-first search,
i.e., $O(b^d)$ with the branching factor $b$ and depth $d$.

ProbLog provides a Python interface which makes it easy to interface with our existing application.
Moreover, Python facilitates data (pre-)processing and enables the execution of arbitrarily complex code.
We therefore parse the substitution results and execute these in Python.

To this end, each relation is linked to an implementation
capturing a string that can be evaluated in Python (Listing~\ref{lst:implementation}).
\begin{lstlisting}[style=problog,
    caption={Python implementation of relations.},
    label=lst:implementation]
implementation(r1, "
dmin.v = min(d_2d.v)
dmin.t = d_2d.t
").
implementation(r2, "
# row width of depth image
w = 320
# take 115th row (about the height of lidar scan)
h = 115
d_2d.v = [d for i, d in enumerate(d_3d.v)
          if i >= h*w and i < (h+1)*w]
d_2d.t = d_3d.t
").
\end{lstlisting}
For instance, \texttt{r1} passes the distance measurements \texttt{d\_2d.v}
through the function it should implement, here: \texttt{min}.
The timestamp \texttt{dmin.t} is adopted from the distance measurements
to indicate the time when the minimum distance occurred.

\section{Fault Detection by Exploiting Redundancy}
\label{sec:monitor}

A generic monitor uses the knowledge base
to setup a monitor for a given signal and to periodically compare against the related signals.

\subsection{Setup and Adaptation}

First, the monitor queries the valid substitutions $S(\var)$
given the variable $\var$ to monitor and the available signals (Sec.~\ref{sec:prolog}).
Then the monitor parses the result and instantiates the substitutions.
Once set up, the monitor periodically executes the substitutions with the given input itoms
and compares the outputs against each other.

On system changes, e.g., regarding the relations or the availability of signals,
the knowledge base can be adapted and the monitor re-initialized.
For instance, the knowledge base may be extended with new relations
whenever new information becomes available
(i.e., when new observation components are connected).

\begin{figure}[htb]
  \centering

\tikzstyle{component} = [draw, fill=black!10, rectangle, rounded corners, inner sep=6pt]
\tikzstyle{io} = [rectangle, inner sep=3pt, align=center, font=\small]

\tikzstyle{var} = [draw, ellipse, inner sep=3pt, align=center]
\tikzstyle{rel} = [draw, rectangle, rounded corners, inner sep=3pt, minimum size = 1.8em]
\tikzstyle{itom} = [rectangle, align=center, font=\small]
\tikzstyle{dummy} = [draw=none]

\begin{tikzpicture}[auto, node distance=1em, >=latex, font=\sffamily\small]

  \matrix (m) [matrix of nodes,
    column sep=1em,
    row sep=1em]{
    $\varp_{sonar}$
    & $\varp_{lidar}$
    & $\varp_{tof}$
    & $\varp_{dmin}$
    \\[1em]
    & 
    & |[rel]| $\{ d_{(x,y,z)} | z = z_{lidar} \}$
    & 
    \\
    |[rel]| $min(.)$
    & |[rel]| $min(.)$
    & |[rel]| $min(.)$
    & 
    \\[2em]
    & 
    & |[component]| compare
    & 
    \\
  };

  \draw [->] (m-1-1) -- (m-3-1);
  \draw [->] (m-3-1) |- ++(0,-8mm) -| (m-4-3.140);

  \draw [->] (m-1-2) -- (m-3-2);
  \draw [->] (m-3-2) |- ++(0,-6mm) -| (m-4-3.120);

  \draw [->] (m-1-3) -- (m-2-3);
  \draw [->] (m-2-3) -- (m-3-3);
  \draw [->] (m-3-3) -- (m-4-3);

  \draw [->] (m-1-4) |- ++(0,-30mm) -| (m-4-3.60);

  \node [below left = 1.7em and 1.5em of m-3-3.south west] (cd1) {};
  \node [below right = 1.7em and 1.5em of m-3-3.south east] (cd2) {};
  \draw [-, dashed] (cd1) -- (cd2);
  \node [right = -0.5em of cd2] (cdtext) {$\var_{dmin}$};

  \node [io, below = of m-4-3] (status) {status};
  \draw [->] (m-4-3) -- (status);

  \node [io, above = of m-1-1] (s0) {$s_0$};
  \node [io, above = of m-1-2] (s1) {$s_1$};
  \node [io, above = of m-1-3] (s2) {$s_2$};
  \node [io, above = of m-1-4] (s3) {$s_3$};

\end{tikzpicture}
  \caption{Check $\varp_{dmin}$ against redundancy.
    The itoms are first transferred into the common domain
    (here: $\var_{dmin}$)
    and compared against each other.}
  \label{fig:monitor}
\end{figure}
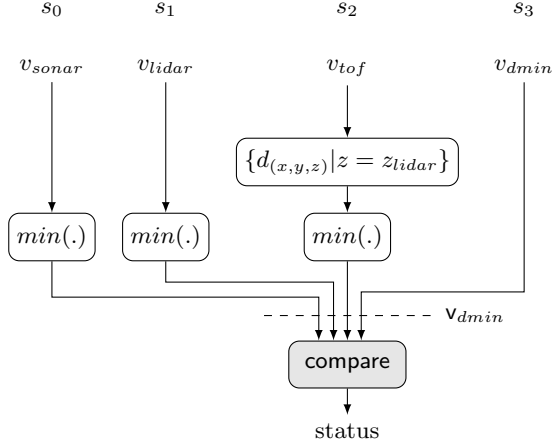

\textbf{Case Study}:
The collision avoidance (controller $z_{safe}$) needs the signal $\varp_{dmin}$
which shall therefore be monitored. The setup is depicted in Figure~\ref{fig:monitor}.
This monitor compares the output of 5 substitutions
(including the empty substitution using the itoms of signal $\varp_{dmin}$ directly).

Note, that the substitutions should be diverse.
For instance, an error in the LIDAR will cause errors in 3 substitutions
including $\varp_{dmin}$ itself ($z_{calc}$ uses $\varp_{lidar}$ as input too).

\subsection{Observation Model}

Once the observations or itoms are in the common domain, their values can be compared to each other.
However, the values are uncertain, e.g., due to noise, interference or environmental conditions
(the accuracy of a LIDARs distance measurement for instance is sensitive to the reflexion capabilities of a surface).

In our proof-of-concept implementation we use simple integer arithmetic
to express the allowed uncertainty of an itom or confidence into an itom.
The value of the itom $\varp(t_s) = (\vinterval{v}, t_s)$
is expressed as an interval~(Eq.~\ref{eq:itom_value}).
\begin{align}
  \label{eq:itom_value}
  \vinterval{v} &= [\vlo{\varp}, \vhi{\varp}] = \{\varp \in \mathbb{R}^2 \, | \, \vlo{\varp} \le x \le \vhi{\varp}\}
\end{align}

In real-world data, however, the itoms diverge not only in space but also in the time domain.
This phenomenon is known as dead-time or aftereffect and occurs in time-delay systems~%
\cite{richard2003timedelay}.
Delays may accumulate from different sources, not only from communication latency.
For instance, the travel of sound is slower than the travel of light.
So the sonar sensor of a mobile robot will react slower
to changes in the distance than a laser scanner.
The time shifts can be estimated, e.g., by using the Skorokhod distance%
~\cite{majumdar2015computing}.
However, the delay of the sonar measurement depends on the distance measured,
i.e., the time delay is not constant.

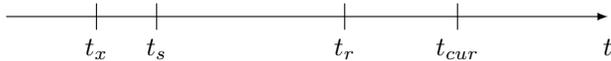
\begin{figure}[h!]
  \centering
  \begin{tikzpicture}[only marks, y=.5cm, >=latex]
  \draw[->] (0,0) -- coordinate (x axis mid) (8,0) coordinate (x axis end);
  \node[below=5pt] at (x axis end) {$t$};

  \draw (1.2,5pt) -- (1.2,-5pt) node[anchor=north] {$t_{x}$};
  \draw (2,5pt) -- (2,-5pt) node[anchor=north] {$t_{s}$};

  \draw (4.5,5pt) -- (4.5,-5pt) node[anchor=north] {$t_{r}$};
  \draw (6,5pt) -- (6,-5pt) node[anchor=north] {$t_{cur}$};
\end{tikzpicture}
  \caption{Timeline of an itom $\varp(t_s)$.}
  \label{fig:timeline}
\end{figure}

We therefore distinguish between
the time $t_x$, when the CPS variable actually adopted the value,
the timestamp of the measurement $t_s$ included in the itom
(typically the time when the message has been generated and provided by the sensor)
and the time the itom is received $t_r$ or used $t_{cur}$ (here: by the monitor),
depicted in Figure~\ref{fig:timeline}.
Unfortunately, $t_x$ is unknown and typically does not match $t_s$
(though often the difference is neglectable).
CPS variables are often sampled periodically.
The sampling period $T$,
i.e., the time between two consecutive itoms of a signal,
should be small enough to follow the dynamics of the CPS variable under observation.

Note that the timeline or the order of these timestamps are applicable in most CPSs,
regardless of the type or implementation of the communication
(e.g., synchronous or asynchronous).
However, the results of the computation using the itom
might get distorted, especially when
$t_s \gg t_x$ (late time-stamping, e.g., due to insufficient sensor capabilities)
or $T < (t_r - t_s)$ (high communication latency, e.g., due to monitoring in the cloud).

For the above reasons, the monitor should not simply compare the last values received
but consider the time delay (from $t_x$ to $t_{cur}$).
Still the itom is the best (latest) the monitor can make use of
(without much more effort like forward estimation).
An itom is therefore considered valid for a \emph{time interval} or time span
(and not only at the given timestamp $t_s$ of the itom).
\begin{align}
  \label{eq:itom}
  \varp(t_s) = (\vinterval{v}, \vinterval{t}) \qquad
  & \vinterval{v} = [\vlo{v}, \vhi{v}] \\
  & \vinterval{t} = [\vlo{t}, \vhi{t}] \quad t_s \in [\vlo{t}, \vhi{t}] \nonumber
\end{align}

An itom is typically late (e.g., acquired by a sensor), but it may be valid/useful into the future,
i.e., the sampling time $t_s$ lies within the time span $\vinterval{t}$.

A variable~$\var$ is \emph{provided} at time $t_{cur}$
when at least one corresponding itom $\vinterval{\varp}(\vinterval{t})$ exists
with $t_{cur} \in \vinterval{t}$.

\begin{figure}[h!]
  \centering
  \tikzstyle{s1} = [mark=*, mark options={color=blue}]
\tikzstyle{s2} = [mark=triangle*, mark options={color=red}]
\tikzstyle{last} = [mark options={color=black}]
\tikzstyle{uncertainty} = [fill opacity=0.5, draw=none]
\tikzstyle{s1_uncertainty} = [uncertainty, fill=blue]
\tikzstyle{s2_uncertainty} = [uncertainty, fill=red]

\begin{tikzpicture}[only marks, y=.5cm]
  \draw[->] (-2pt,0) -- coordinate (x axis mid) (7.5,0);
  \node[below=3pt] at (x axis mid) {$t$};

  \draw[->] (0,-2pt) -- coordinate (y axis mid) (0,7);
  \node[left=3pt] at (y axis mid) {$\varp(t)$};

  \foreach \t in {1,2,3,4,5,6}
  \draw[s1] plot coordinates {(\t,{1+4*sin(\t*20)})};

  \draw[dotted] (6,4.5) -- (6,2) node[anchor=north] {$t_{s}$};
  \draw[s1_uncertainty] (5,{1+4*sin(6*20)-0.2}) rectangle (6,{1+4*sin(6*20)+0.2});
  \draw[s1,last] plot coordinates {(6,{1+4*sin(6*20)})};
  \draw[<->] (5,{1+4*sin(6*20)-0.5}) -- node[below] {$\Delta = T$} (6,{1+4*sin(6*20)-0.5});

  \foreach \t in {0.8,1.8,2.8,3.8,4.8,5.8}
  \draw[s2] plot coordinates {(\t,{1.5+4*sin(\t*20)})};

  \draw[s2_uncertainty] (4.8,{1.5+4*sin(5.8*20)-0.6}) rectangle (5.8,{1.5+4*sin(5.8*20)+0.6});
  \draw[s2,last] plot coordinates {(5.8,{1.5+4*sin(5.8*20)})};

  \draw[dotted] (6.2,6.5) -- (6.2,-3pt) node[anchor=north] {$t_{cur}$};
\end{tikzpicture}
  \caption{Two one-dimensional signals (itoms over time).
    The confidence regions of the latest itoms are highlighted.}
  \label{fig:signal}
\end{figure}
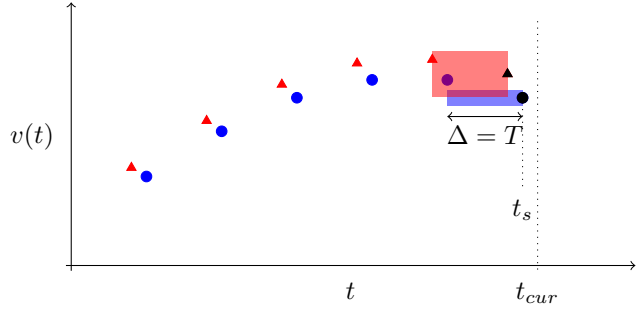

\textbf{Case Study}:
We set $\vinterval{t} = [t_s - \Delta, t_s]$
given the timestamp of the observation $t_s$ and a delay~$\Delta$
which sums up the latencies caused by the sampling process, pre-processing (e.g., filtering) and communication.
The size of the interval may match the sampling period~$T$
of the signal to continuously have comparable values.
Note, that a sensor should be able to follow the dynamics of the system ($T$ is small enough).
Figure~\ref{fig:signal} visualizes two signals of the same variable
and the confidence intervals of the latest itoms received by the monitor.
For a one-dimensional itom the confidence region is a rectangle.

The demonstration in Section~\ref{sec:experiments}
uses the specifications from sensors' datasheets.
For instance, we set the maximum delay $\Delta_{sonar}$
of the itoms provided by the array of $8$ sonar sensors to $8 \cdot 10ms$,
considering the maximum range of $3m$, the sonar speed of $343m/s$
and an average communication delay of $1ms$.

\subsection{Comparison}

Exceeding the threshold for an error is implicitly defined by the size of the intervals (in time and space).
Overlapping confidence intervals indicate the itoms to be equal (Fig.~\ref{fig:signal}),
non-overlapping means the itoms diverge.
It is a relative fast way to compare itoms (cf. itoms represented by a probability distribution),
and therefore suitable for real-time systems.

The monitor is periodically ($T_m$) fed with the latest itoms.
It may save previous itoms to compensate for late receptions,
so a newly received, but delayed itom ($\vinterval{t} \ll t_{r} < t_{cur}$) can be compared to a past itom with similar time interval.

The total set of buffered itoms $I$ is used as inputs to the substitutions $S(\var)$
to generate comparable (output) itoms.
Two or more itoms are \emph{comparable}
when the itoms have a common corresponding variable and their time intervals overlap.

A monitor step consists of the following substeps:
\begin{itemize}
  \item
    \emph{Collect combinations}
    of input itoms~$(\vinterval{v},\vinterval{t})$
    per substitution~$s \in S(\var)$ such that their time intervals overlap
    and a combination $c$ contains exactly one itom per input signal $\varp \in I_s$ of $s$
    (Cartesian product).
\end{itemize}
\begin{align*}
  & \forall s \in S(\var) , \, v_1,..,v_m \in I_s\colon \\
  & C_s = \{ ((\vinterval{v},\vinterval{t})_1 , \hdots , (\vinterval{v},\vinterval{t})_m)
  \, | \, \cap_{i=1,..,m} \vinterval{t}_{i} \ne \emptyset \}
\end{align*}

\begin{itemize}
  \item
    \emph{Execute the substitutions}
    to generate output itoms to compare against.
The value interval of the input itoms~$c \in C_s$ of a substitution $s$
are passed through the functions of $s$
giving the value interval of the output itom
(see interval arithmetic~\cite{hickey2001interval}).
The time interval of the output itom is the intersection of the time intervals of the inputs.
\end{itemize}
\begin{align*}
  & \forall s \in S(\var) , \, \forall c \in C_s\colon \\
  & O = \{ (s, (\vinterval{v}, \vinterval{t})) \, | \,
  \vinterval{v} = \vinterval{f}_s(c) , \,
  \vinterval{t} = \bigcap_{\varp_i \in c} \vinterval{t}_i
  \}
\end{align*}

\begin{itemize}
  \item
    \emph{Pairwise compare}
    the output itoms $O$ to calculate an error matrix of the value intervals.
\end{itemize}
\begin{align*}
  & \forall (s,(\vinterval{v},\vinterval{t}))_i,
  (s,(\vinterval{v},\vinterval{t}))_j \in O , \, i \ne j\colon \\
  & e_{s_is_j} = \sum |max(\vlo{v}_i, \vlo{v}_j) - min(\vhi{v}_i, \vhi{v}_j)|
\end{align*}

\begin{itemize}
  \item
    \emph{Rank the error}
    to return the substitution with the highest error.
    The errors per substitution are summed up (sum of columns or rows of the error matrix).
    When an error per substitution is greater than $0$ the monitor returns the failed substitution.
\end{itemize}

\textbf{Case Study}:
The implementation uses pyinterval~\footnote{\url{https://pypi.org/project/pyinterval/}}
to pass itoms through a substitution
or to identify the pairs of itoms to compare.
The size of the itoms buffer is a constant parameter of the monitor for simplicity.
However, the monitor may increase the size of the buffer
according to the observed delays ($t_r - t_s$).

\section{Experiments}
\label{sec:experiments}

In this section we demonstrate the monitor setup and fault detection.

\subsection{Setup}

The demo uses the mobile robot described in Sec.~\ref{sec:background}.
The hosts on the rover (connected via WiFi to the notebook)
only run the sensor drivers,
while the controllers and the monitor are executed on the notebook
(Intel i7 2.1GHz x 4, 8GB RAM, Ubuntu, ROS and its nodes run in Docker containers).

\begin{figure}[htb]
  \begin{subfigure}[b]{0.49\linewidth}
    \includegraphics[width=\linewidth]{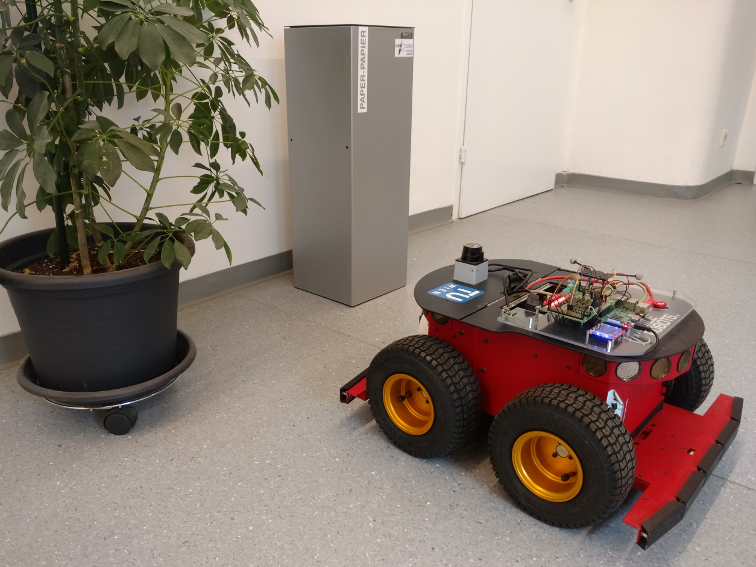}
  \end{subfigure}
  \hfill
  \begin{subfigure}[b]{0.49\linewidth}
    \includegraphics[width=\linewidth]{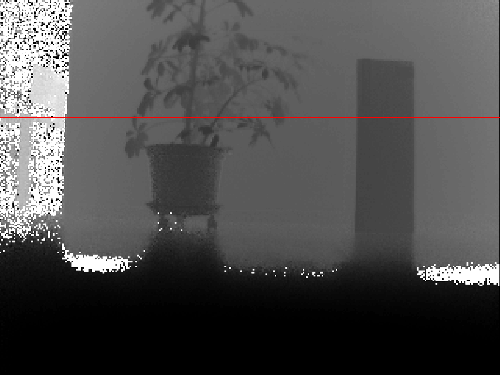}
  \end{subfigure}
  \caption{Images of the test scenario.
    The figure on the right depicts a ToF camera depth image.
    The line marked in red is used to compare against the laser scan.}
  \label{fig:scenario}
\end{figure}

The rover is tele-operated via the notebook towards a wall
(Fig.~\ref{fig:scenario}).
%
The ROS messages of the sensors (monitored inputs)
and the monitor itself (failed substitution and additional debug output)
are logged.

\subsection{Demonstration}

\begin{figure}[t]
  \centering
  \includegraphics[width=\linewidth]{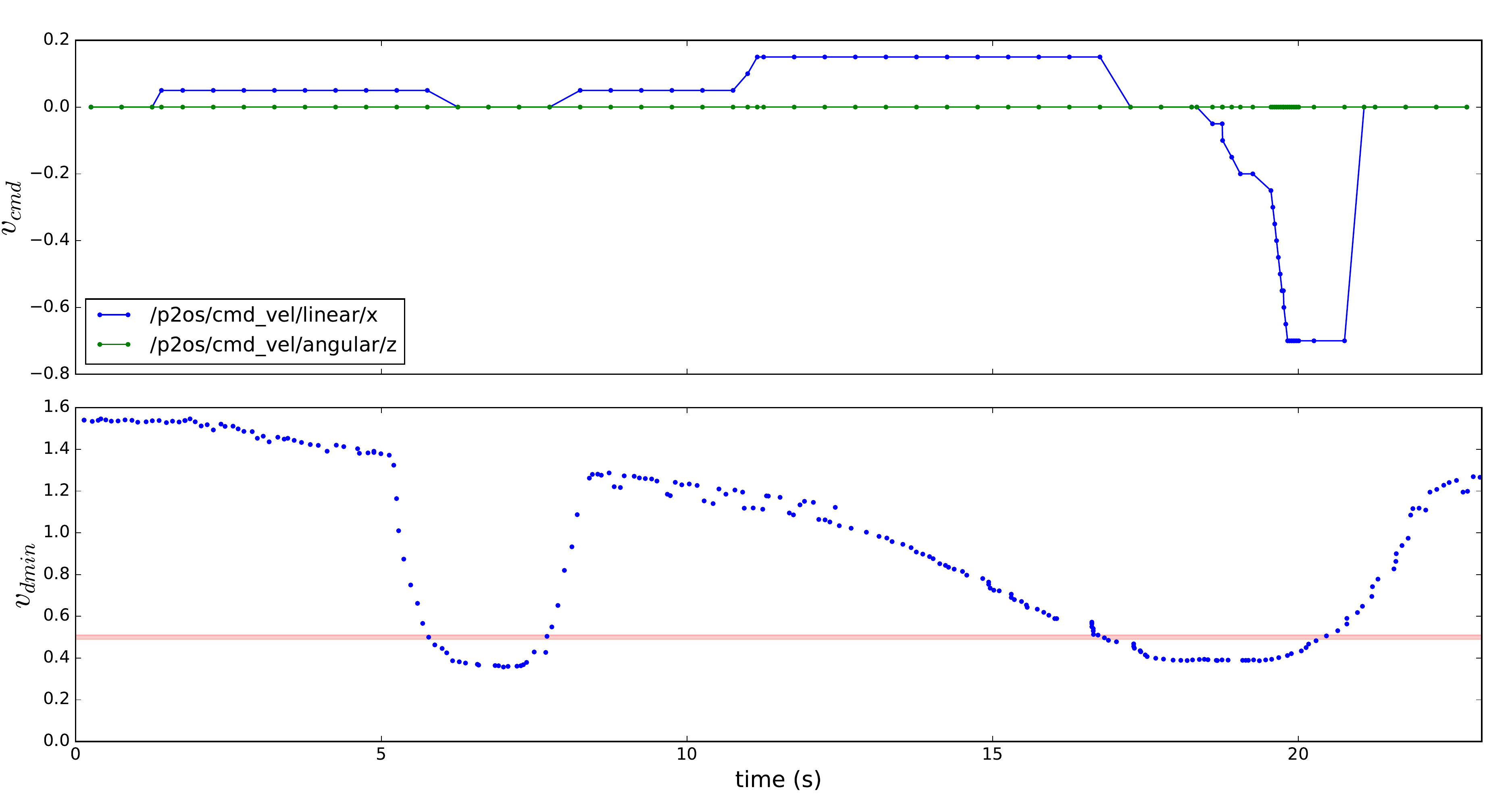}
  \caption{Visualization of an application run in the scenario.
    Plots (from top to bottom):
    (1)~speed commands (linear and angular velocity) from the notebook,
    (2)~minimum distance to an obstacle $\varp_{dmin}$ (based on LIDAR data)
    - the red area indicates the collision avoidance threshold.}
  \label{fig:demolog}
\end{figure}

Figure~\ref{fig:demolog} depicts some application logs taken from a run of the demo.
The rover first moves slowly and straight towards the wall
($t = 2..5s$: $v_{dmin}$ slowly decreases).
To test the possibly different delays of the sensors,
a dynamic obstacle has been placed in front of the rover for some period of time
($t = 5..8s$: $v_{dmin}$ falls below a threshold, robot stops).
Finally, the rover reaches the wall ($t = 17..20s$)
and goes back relatively fast ($t = 20..23s$).

\begin{figure}[t]
  \centering
  \includegraphics[width=\linewidth]{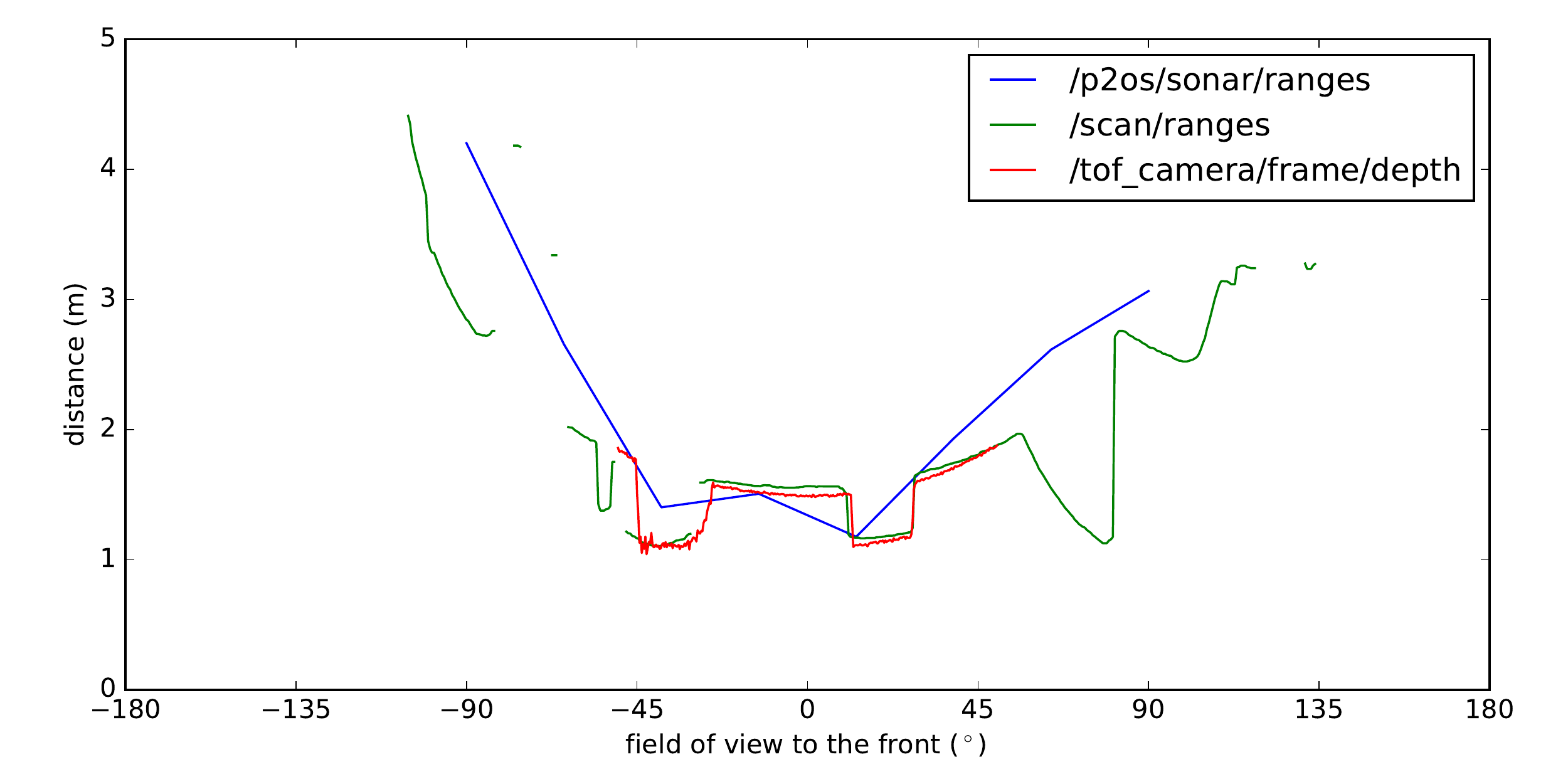}
  \caption{Signals in the domain $\var_{d\_2d}$ at about 7s.
    Observations of the sensors measuring distance to obstacles (sonar, LIDAR, ToF camera).}
  \label{fig:measurements}
\end{figure}
The distance measurements in Figure~\ref{fig:measurements}) show the significant differences,
e.g., in field of view (the smallest is the one of the ToF camera) or
resolution (8 measurements from the sonar vs. 320 from the ToF camera vs. 512 from the LIDAR),
of the 3 sensors: sonar array, LIDAR and ToF camera.
However, in the domain of $\var_{dmin}$ the observations can be reasonably compared
though outliers have to be reckoned.

\begin{figure}[htb]
  \centering
  \includegraphics[width=\linewidth]{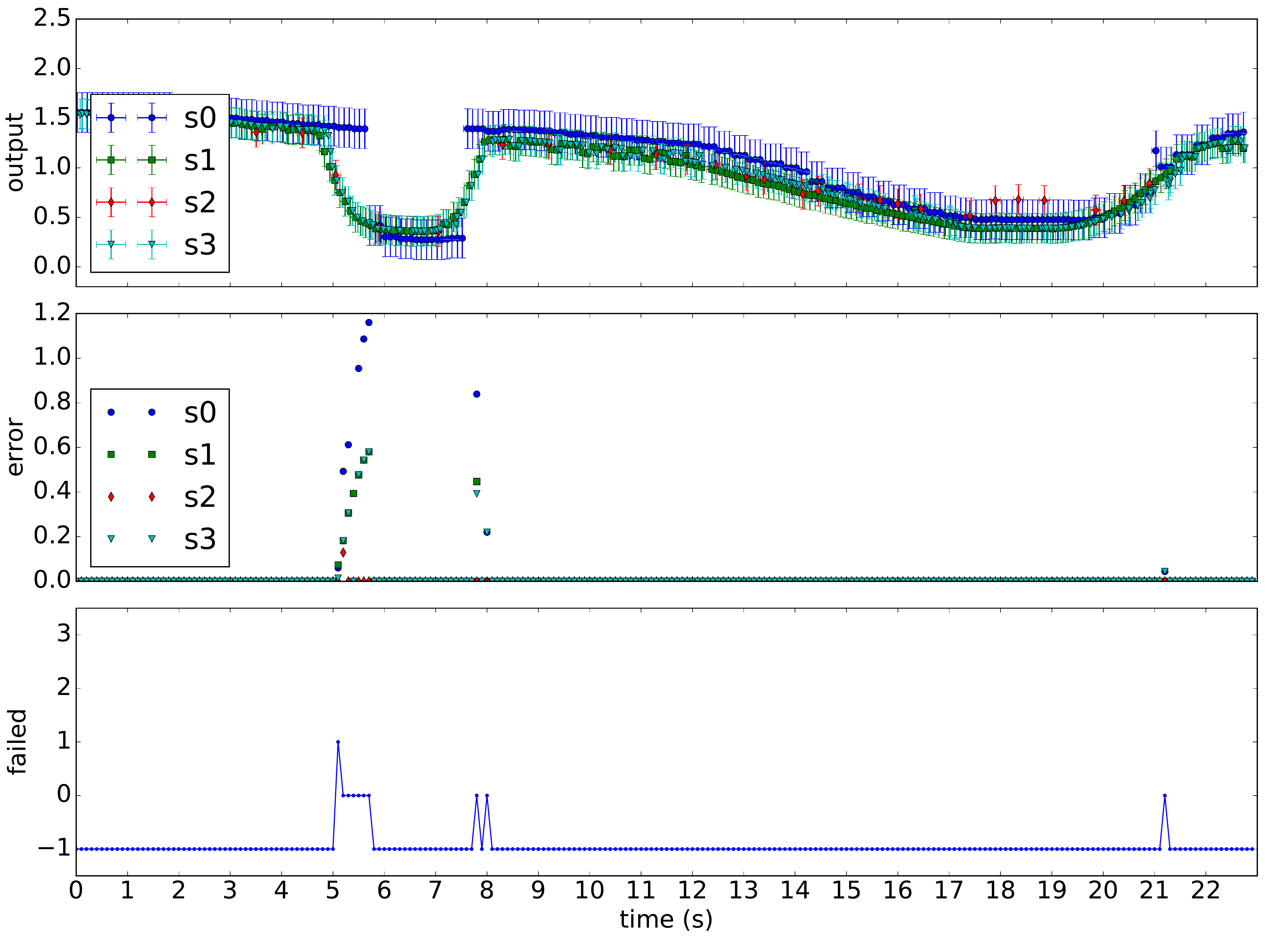}
  \caption{Monitor outputs over time of the demonstration.
    Plots (from top to bottom):
    (1)~$\var_{dmin}$, outputs of the substitutions (executed by the monitor),
    (2)~error per substitution (calculated by the monitor),
    (3)~output status of the monitor -- index of the failed substitution or -1 if all comparable substitution outputs match.}
  \label{fig:monitorlog}
\end{figure}

Figure~\ref{fig:monitorlog} runs the monitor on the itoms collected above.
The monitor triggers some faults due to improper timestamps of the sonar array
(late when obstacle appears, but too early when the obstacle disappears)
and an outlier of the same.




\subsection{Fault Detection}

To visualize and test the fault detection
we inject faults to a generated set of itoms
The itoms per monitor step have equal value and slightly shifted timestamps.

Table~\ref{tab:fault-injection} summarizes the types of faults,
how these are injected and the expected output of the monitor.
Note that transient faults (outliers) can be mitigated by
the additionally provided moving-median or moving-average filter of the monitor output.
We therefore focus on faults occurring over a longer period of time (or permanent faults)
and turn the filters off for reasonable plots of the results.
All faults are injected for a specific period of time $[t_{start},t_{end}]$.
$k$ faults can be identified when the number of comparable outputs is $\ge 2k+1$
(for Byzantine faults: $3k+1$).
To be able to identify a fault in our setup, we assume only one simultaneous fault in the value domain.
The buffer size $n_{buf}$ of the monitor is set to 1 for testing faults in the value domain.
\newlength{\faulttabwidth}
\faulttabwidth0.9\linewidth
\begin{table}[t]
  \centering
  \begin{tabular}{|p{0.2\faulttabwidth}|p{0.8\faulttabwidth}|}
    \hline
    \textsc{Fault} & \textsc{Injection}
    \\ \hline
    \hline
    Random noise
    & Add noise sample $x \sim U(a,b)$ to the observation.
    \\ \hline
    Stuck-at 0
    & Set observation to $0$.
    \\ \hline
    Time shift
    & Add the duration $\Delta$ to the \emph{reception} timestamp $t_r$
    of the observation.
    \\ \hline
    Missing or lost messages
    & Delay observations by $\Delta > n_{buf} \cdot T_{m}$ (implemented by time shift).
    \\ \hline
  \end{tabular}%
  \caption{Fault injection to an input signal of a substitution.}
  \label{tab:fault-injection}
\end{table}

\begin{figure}[t]
  \centering
  \includegraphics[width=\linewidth]{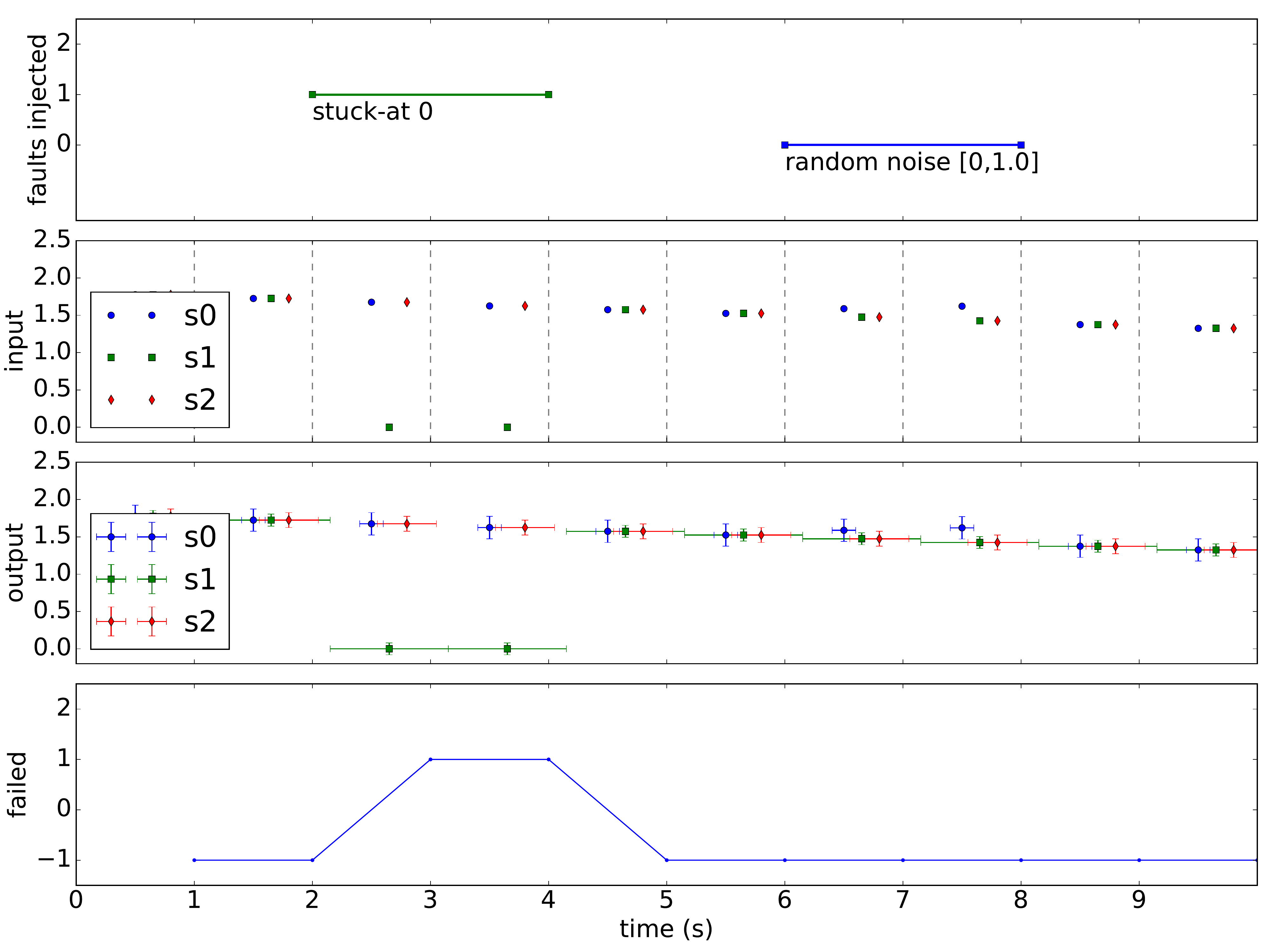}
  \caption{Fault detection in the value domain.
    From top to bottom:
    (1)~reception of inputs (points) and monitor calls (dashed vertical lines),
    (2)~outputs of the substitutions including uncertainty,
    (3)~fault injection,
    (4)~monitor output (index of the failed substitution or -1 if all substitution outputs match).}
  \label{fig:faultlog_value}
\end{figure}

\begin{figure}[t]
  \centering
  \includegraphics[width=\linewidth]{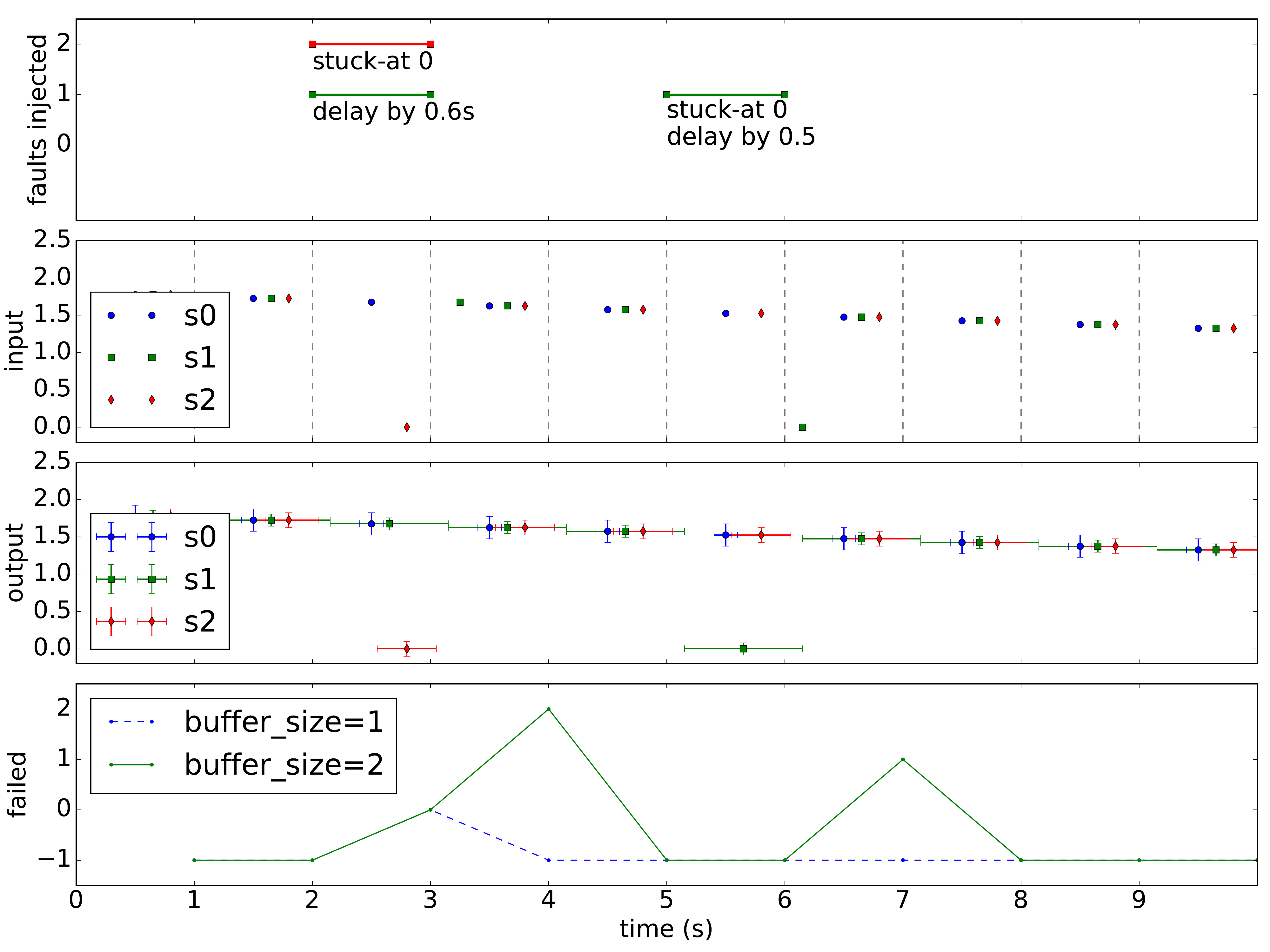}
  \caption{Fault detection in the time domain.
    From top to bottom:
    (1)~fault injection,
    (2)~reception of inputs (points) and monitor calls (dashed vertical lines),
    (3)~outputs of the substitutions of the monitor $n_{buf}$=2,
    (4)~monitor output (index of the failed substitution or -1 if all substitution outputs match).}
  \label{fig:faultlog_time}
\end{figure}

Figures~\ref{fig:faultlog_value} and \ref{fig:faultlog_time} visualizes
fault detection of 3 substitution outputs.
The top plot annotates the time span with a description of the injected fault.
The height of the bar references the substitution index to which the fault has been injected.
The generated value is represented by a marker
while the uncertainty (intervals) is plotted as two error bars
along the x- (time) and y-axis (value).
The upper plot presents the timeline of reception of the monitor inputs
($t_r$ of the itoms).
The lower plot shows the outputs of the substitution executed by the monitor,
i.e., the itoms at the time of occurrence
($t_s$ or $\vinterval{t}$ of the itoms).
The bottom indicates the substitution index with the highest error.

For instance, in Fig.~\ref{fig:faultlog_value} stuck-at 0 and random noise
has been applied to the input of substitution $s_1$ and $s_0$, respectively.
The monitor gathers the itoms since the last call (here within the last second)
and compares the outputs of the substitutions when the time intervals overlap.
Because the value intervals of $s_1$ during the time marked with ``stuck-at 0''
do not overlap the value intervals of the other substitutions
$s_1$ has an error $\ne 0$ and the monitor indicates the failed substitution as $s_1$.

In Fig.~\ref{fig:faultlog_time} the capability to compensate delays is tested.
The input of $s_1$ is missing in the monitor call at time $3s$.
The faulty substitution is miss-classified because of only two available outputs to compare
(no majority vote possible).
However, with a buffer size of $n_{buf} = 2$
the fault can be detected
(the monitor collects itoms within $n_{buf} \cdot T_m$, here, the last 2 monitor periods).
The delayed itom of $s_1$ can be compared to the itoms of the last call.
It matches the output of $s_0$, so the monitor classifys $s_2$ as faulty at time $4s$.

The logged data and the scripts to inject faults and run the monitor offline
can be found on GitHub~\footnote{\url{https://github.com/tuw-cpsg/paper-shsa-monitor-experiments}}.

\section{Conclusion}
\label{sec:conclusion}

A cyber-physical system (CPS) assembled out of many sub-systems
provides observations of different CPS variables
which can be interrelated to each other in a knowledge base.
We present a monitor querying the knowledge base to find redundancies
and which can detect faults of observation components by comparison.
The knowledge base has been encoded in Prolog
implemented with the ProbLog library
which enables the user to change the knowledge base during runtime,
and add or remove information about the availability of observations.
The monitor can therefore master technological or functional changes in the CPS.
Moreover, we presented an observation model
considering uncertainties in space and time (e.g., noise or delays)
of observations collected by the monitor.

In ongoing and future work we want to investigate other methods
to compare and rank the redundant observations over time.
Furthermore, we look into how the redundancy shall be selected (diversity)
or proper fault diagnosis can be applied.

\section*{Acknowledgment}
The authors thank Peter Puschner for proof-reading the paper.
The research leading to these results has received funding from the IoT4CPS project
partially funded by the ``ICT of the Future'' Program of the FFG and the BMVIT,
as well as from the Productive 4.0 project
co-funded by the European Union and ECSEL.

\bibliographystyle{unsrt}
\bibliography{bib/references}

\end{document}